\documentclass[Afour,sageh,times]{sagej}

\usepackage{moreverb,url}
\usepackage{natbib}
\usepackage{multicol}

\usepackage[colorlinks,bookmarksopen,bookmarksnumbered,citecolor=red,urlcolor=red]{hyperref}

\newcommand\BibTeX{{\rmfamily B\kern-.05em \textsc{i\kern-.025em b}\kern-.08em
T\kern-.1667em\lower.7ex\hbox{E}\kern-.125emX}}

\def\volumeyear{2016}

\begin{document}

%\runninghead{Ng et. al}

\title{Cyborgs for strategic communication on social media}

\author{Lynnette Hui Xian Ng\affilnum{1}, Dawn C. Robertson\affilnum{1} and Kathleen M. Carley\affilnum{1}}

\affiliation{\affilnum{1}Carnegie Mellon University, Pittsburgh, USA}

\corrauth{Lynnette Hui Xian Ng, Computational Analysis of Social and Organizational Systems, Carnegie Mellon University, Pittsburgh, USA}

\email{lynnetteng@cmu.edu}

%\author{Anonymous Submission}

\begin{abstract}
Social media platforms are a key ground of information consumption and dissemination. Key figures like politicians, celebrities and activists have leveraged on its wide user base for strategic communication. Strategic communications, or StratCom, is the deliberate act of information creation and distribution. Its techniques are used by these key figures for establishing their brand and amplifying their messages. Automated scripts are used on top of personal touches to quickly and effectively perform these tasks. The combination of automation and manual online posting creates a Cyborg social media profile, which is a hybrid between bot and human. In this study, we establish a quantitative definition for a Cyborg account, which is an account that are detected as bots in one time window, and identified as humans in another. This definition makes use of frequent changes of bot classification labels and large differences in bot likelihood scores to identify Cyborgs. We perform a large-scale analysis across over 3.1 million users from Twitter collected from two key events, the 2020 Coronavirus pandemic and 2020 US Elections. We extract Cyborgs from two datasets and employ tools from network science, natural language processing and manual annotation to characterize Cyborg accounts. Our analyses identify Cyborg accounts are mostly constructed for strategic communication uses, have a strong duality in their bot/human classification and are tactically positioned in the social media network, aiding these accounts to promote their desired content. Cyborgs are also discovered to have long online lives, indicating their ability to evade bot detectors, or the graciousness of platforms to allow their operations.
\end{abstract}


\maketitle

\section*{Introduction}
Cyborgs are social media accounts that are not always bots, yet are not always people. They are a hybrid of both worlds: accounts that are detected as bots in one time window, but identified as humans in another.  They present this outfit because they are controlled by human operators at some instances, yet controlled by automated scripts in other instances \citep{gorwa2020unpacking}, resulting in differing bot classification, dependent on the point of measurement.

Strategic communications (StratCom) is the measured act of creating and pushing information to the public \citep{hallahan2007defining}. The potential for quick and vast information dissemination makes social media an ideal candidate for StratCom. Public personas and organizations use StratCom techniques on social media as part of their strategy to establish brand image, build consensus on important issues and market their products \citep{borchers2019social,hallahan2007defining}. Countries also employ StratCom techniques in social media during Russia-Ukraine conflict as early as 2015, building and communicating narratives to different audiences \citep{lange2015strategic}.

Past research of Cyborgs involves the identification of these semi-automated accounts through machine learning techniques that separate feature spaces like linguistic cues (i.e., number of pronouns, number of hashtags) or account meta-data (i.e., number of followers) \citep{chu2012detecting,castillo2019detection}. However, not only do these supervised learning techniques require data where agents are pre-annotated for being Cyborgs, but they also do not exploit the fundamental definition of Cyborgs -- their semi-bot-semi-human duality. \citet{castillo2019detection} further revealed that automatic agent classification using such feature-based machine learning methods perform with a mediocre accuracy of 60\%, indicating that there is a need for other types of methods to differentiate Cyborgs. More generally, previous work do not study the nature of Cyborg accounts and their primary communicative use, especially in the complex social network environment.

We begin by theorizing a quantitative definition for Cyborgs. Cyborgs can be definitively detected as bots or as humans in different timeframes. This definition provides us with two properties that can be measured quantitatively: (1) the changes in bot classification outputs that changes the classification of bot/human of the Cyborg agent from timeframe to timeframe; and (2) the difference in bot-likeness scores that provides definitive bot/human classification. Therefore, quantitatively, Cyborgs can be measured with bot detection algorithms through: (1) frequent flipping of bot classification of the agent, thus changing their bot/human labels from timeframe to timeframe; and (2) large difference in bot likelihood scores between the flips, stipulating a definitive change in behavior rather than a tremor in conduct. Establishing a quantitative definition for Cyborgs provides a repeatable methodology to identify these agents rather than rely on subjective judgment.

This work sheds light on the use of Cyborgs in the context of two key events: the 2020 Coronavirus pandemic and the 2020 US Elections, covering over 63 million agents. These two events were widely covered by media outlets, involved several influential personas and sparked many contentious issues. With data collected across two years, we analyzed social media agents on Twitter that exhibit strong Cyborg behavior in their inconsistent bot classification. These accounts were observed to be used primarily for Strategic Communications (StratCom) for public figures and activists. Ultimately, our analyses aim to identify social media agents that are constructed for strategic communication uses, characterize the duality of their bot classification and their tactical social network positions. The study's research questions are summarized as follows: 
\begin{enumerate}
    \item \textbf{RQ1:} What are suitable threshold values for identifying Cyborgs? Through the use of longitudinal time series analysis, we establish suitable threshold values for the quantitative definition of Cyborgs: the number of times for an agent to flip bot classification and the average difference in bot probability score between flips to be considered a Cyborg.
    \item \textbf{RQ2:} What are the differences between Cyborgs and Non-Cyborgs in terms of their social network positions and discourse topics? Through network centrality metrics and topic analysis comparison, we establish the general differences of the two types of agents within the collected dataset.
    \item \textbf{RQ3:} What are Cyborgs used for? Through manual annotation, we identified that Cyborgs are used for strategic communications, having a mixture of automated posts and human-written posts. 
\end{enumerate}

Throughout this paper, since a large proportion of identified users are active accounts we cannot share their account handles nor screenshots of their pages to preserver user privacy. We describe users in general and do not disclose specific identifiers.

\section{Related Work}
\subsection{Strategic Communications}
Strategic Communications, or StratCom for short, is purposeful communication by individuals or organizations, for brand promotion and relationship building. It is used in several domains, including management, marketing, public relations, political communication and information campaigns \citep{hallahan2007defining}. Groups of people that engage in the deliberate development, curation and dissemination of information include governments, companies, activist organizations, and leaders of government, corporations or special interest groups. StratCom sets itself apart from general communicative discourse because the medium and audiences are not simply channels of communication and message receivers, but its practitioners consciously use the media to shape social and cultural realities, such as promoting a leader's brand or championing a social course \citep{holtzhausen2014strategic}.

While StratCom in the past had mostly focused on print and mainstream media, a large portion of institutions have since included all forms of the Internet and electronic communication \citep{holtzhausen2014strategic}. The Internet, and in particular, social media, has provided the ability to disseminate information at high speed and little cost, and have been a helping tool for StratCom. 

One such use of StratCom techniques on the Internet is the use of messaging techniques on social media during the 2014 Russia-Ukraine conflict. Through the use of coordinated narratives, pro-Russian social media agents systematically cultivated fear, anxiety and hate among the Russian population towards the Ukraine population. This was done by posting images, videos and pleas for help of Ukraine atrocities and violence towards Russians on social media sites like Vkontakte and Twitter. To discredit the Western policy, Russian communication released an intercepted phone call between the US Assistant Secretary of State Victoria Nuland and US Ambassador to Ukraine Geoffrey Pyatt over Twitter and YouTube. This leaked phone call also served to indicate that the Russians have access to the US communication lines \citep{lange2015strategic}. 

In today's information-laden world, StratCom is particularly important to organizations to vie for the attention, alignment and allegiance of constituents. In the online space, the alignment of a user is represented as the expressed stance of the social media users, which can be determined through their posts. Stance detection is the computational task to measure the alignment of the author of a text towards a proposition, such as in support of vaccination (pro-vaccine) or against vaccination (anti-vaccine). Stance detection has been used to determine opposing groups in debates, social and political opinion \citep{rajadesingan2014identifying,ng2022pro,sobhani2017dataset}. This task is typically a supervised learning task, in which machine learning models are trained with manually annotated data to differentiate between stances. These models range from support vector machines \citep{elfardy2016cu}, to logistic regressions \citep{augenstein2016usfd} to neural network models \citep{wei2016pkudblab,kawintiranon2021knowledge}.

In this work, we made use of stance detection to evaluate whether there is a difference in terms of the expressed stances from strategic communications and general discourse.

\subsection{Cyborgs}
Social media Cyborgs have been established as agents, or users, that are partially bot and partially human. Several definitions of the term ``Cyborgs" within the social media space have been proposed. \citep{orabi2020detection} and \citep{chu2012detecting} focus on the actor behind the account, defining Cyborgs as ``Human accounts that use automation techniques or bot accounts managed by human beings". \citet{gorwa2020unpacking} states that Cyborgs are hybrid accounts, that ``exhibit a combination of automation and human curation". The Cyborg is also mentioned as an ambiguous account which have mixed human and bot behavior \citep{alarifi2016twitter}.

Overall, definitions of Cyborgs all rely on the dual-nature of the social media account, that they sometimes appear as a bot, and other times behave as a human. \citet{chu2012detecting} defines this duality as either bot-assisted human or human-assisted bot. That is, these accounts are humans with periodic automated assistance, or automated accounts with occasional human intervention. 

Hybrid Cyborg accounts have been annotated and differentiated from real accounts by user profile analysis, temporal analysis, linguistic analysis and social-context analysis \citep{chu2012detecting,orabi2020detection}. \citet{alarifi2016twitter} created a Chrome browser extension that indicates whether a Twitter user was a Cyborg based on a set of extracted features such as number of hashtags per tweet, number of times the user has been retweeted and so forth. \citet{igawa2016account} exploited pattern recognition through Discrete Wavelet Transform of the texts in an agent's posts for Cyborg classification. A similar suite of extracted account characteristics were used by \citet{castillo2019detection} to analyze bots and cyborgs during the 2017 Chilean presidential election, and identify social media accounts of presidential candidates and groups of affiliated users that appear as Cyborgs in their political strategic communications. These are observed through the periodicity of the posts and the usage of third-party applications (e.g. TweetDeck) for some posts.

Cyborgs have also been described as poorly understood agents, for it is unsure how much the alternation between automation and human intervention is required to make a bot a Cyborg \citep{gorwa2020unpacking}. While the dual-nature of Cyborgs have been well-documented, they have not been analytically quantified. This paper contributes to the Cyborg literature by empirically determining the characteristics of social media Cyborgs through exploiting the dual nature of Cyborg accounts, and analyze their social network behavior and communication usage.

\section{Data and Methodology}
\subsection{Data}
In this work, we identified and evaluated Cyborgs from the social media platform Twitter, across two large datasets relating to the Coronavirus pandemic and the US Elections. We collected Twitter data through Twitter's V1 streaming API by retrieving all messages that contained at least one relevant hashtag. We performed this collection across two major events in 2020: the Coronavirus pandemic and the US Elections. The 2020 coronavirus pandemic is a global virus outbreak caused by the virus SARS-CoV-2. Governments and public health authorities imposed restraints such as requiring masks and lockdowns. The 2020 US Presidential Elections was held on November 3 2020, where Democratic presidential nominee Joe Biden won against incumbent Republican Donald Trump. Since the social media discourse was extremely voluminous during the events, we performed this collection every two months. We also filtered the collection stream to only return English tweets.

For the Coronavirus pandemic, we used the keywords related to \#coronavirus to stream for English tweets during June 2020 to May 2021. For the US Elections, we used the keywords related to \#uselections2020 to stream for English tweets during April 2020 to February 2021. The full list of hashtags used is reflected in the Supplementary Material.

\begin{table*}\centering
\caption{Statistics of Data Collection}
\label{tab:statistics}
\begin{tabular}{rrr}
 & Coronavirus dataset & US Elections dataset \\
\midrule
Total agents & 62,072,853 & 23,933,084 \\ 
Total tweets & 355,743,163 & 193,821,760\\ 
Number of Consistent agents & 2,251,974 & 934,978 \\
Number of Cyborgs (\%) & 357,031 (15.85\%) & 199,674 (21.36\%) \\
\bottomrule
\end{tabular}
\end{table*}

In total, there were 62,072,853 agents with 355,743,163 tweets collected for the coronavirus dataset; and 23,933,084 agents with 193,821,760 tweets collected for the US Election dataset. For consistency, we extracted the agents that were consistently present throughout all the months of collection. This provides for a coherent comparison throughout the entire study. In total, we study 3.1 million agents.
\autoref{tab:statistics} presents the statistics of the data used in this paper.

\subsection{Identifying Cyborgs}
The key characteristic of a Cyborg is that it is sometimes a bot, and sometimes a human. We focus on this dual behavior of the Cyborg to establish a definition of a Cyborg based on quantitative thresholds. In our work, a Cyborg is defined as a social media agent that has a frequently flip bot classification, and a large change in bot probability scores between those flips. Therefore, we harness bot classification algorithms to segregate the time periods where an agent is classified as a bot and a human. Many bot classification algorithms have been developed to identify autonomous agents \citep{feng2022twibot,ng2023botbuster}. We select to use the BotHunter classification model \citep{beskow2018bot} to assign a bot probability score that ranges between 0 and 1 to each agent within the datasets. This algorithm is selected because performs the classification evaluation in a local setting, allowing the use of historical tweets, and also evaluate an agent's classification from segregated subsets of data. This bot detection algorithm takes in account post periodicity and source (i.e. Twitter for Android, TweetDeck), which has been shown to be indications of Cyborg functionality \citep{castillo2019detection}. We determined an agent to be a bot when its bot probability is greater or equal to 0.70, and a human otherwise. This value is referenced from previous systematic studies of bot classification models \citep{ng2022stabilizing,rauchfleisch2020false}. Using this value, we traced the number of times an account flips bot classification across the months by measuring the difference in bot classification between each day.  Further details are described in the Supplementary material.

%add botbuster to that previous studies citation ^

Having annotated a bot probability score for each agent for each day, we are able to compare changes in bot probability scores across time. Using the changes in bot scores across time, we identify the number of changes of bot classification per agent, and the difference in bot scores during the changes. Cyborgs are thus agents that have a high number of changes of bot classification, with a high average difference of bot scores during the change. We use the values where the proportion of users that change classification tapers off as our quantitative threshold values for identifying Cyborgs. This is at the 75th-percentile of the overall distribution, with agents having a minimum of 3 flips. We then extract these agents that highly exhibit the Cyborg property.

\subsection{Network Analysis of Cyborgs}
Cyborgs work within the social media ecosystem, and hence interact with other agents in the system. To understand the structure of the networks in which Cyborgs live and operate in, and the influence Cyborgs have on their surrounding neighbors, we analyze social media network metrics. This analysis provides ideas to the patterns of relationships among individuals in the network.

For analyzing the network metrics characterized by Cyborg and non-Cyborg agents, we form an all-communication network for each sub-dataset. This network links two agents if they make a communication interaction with each other. Communication interactions include retweet, quote tweet and @mentions. We then evaluate network centrality metrics of the agents within this network, which indicates the influence of the agent within the network. These network centrality metrics are calculated using the software ORA\footnote{ORA is available from \url{http://www.casos.cs.cmu.edu/projects/ora/}}. The software reads in an XML file that represents a graph of the network, in which the agents are nodes and the communication interactions are links. It then performs the necessary mathematical calculations to output the network centrality values. We selected the ORA software as it facilitates reading in of tweets directly from the collected format of the Twitter API to construct the all-communication network and calculate the network metrics within the software.

Specifically, we analyzed the following network centrality values: betweenness centrality, eigenvector centrality and total degree centrality. Betweenness centrality measures how much an agent lies on a path between two nodes, or groups of nodes. An agent with high betweenness centrality acts as an information broker, because information transverses through it from one agent to another. Eigenvector centrality measures how much influence an agent has, with the virtue of how connected it is to other influential agents. An agent that is more connected to other influential agent has a higher eigenvector centrality value. Total degree centrality is a measure of how many links an agent has, i.e. how many other agents the agent is connect to. The more links an agent has, the more communication has transpired between the agent and other agents. 

In addition to centrality values, we also compare the metadata information of the agents, specifically the number of followers and friends that they have. These information give us an idea of how popular and influential the agents are within their direct social network. The number of followers indicates the extent of influence the agent posses, and the number of friends indicate the extent of reciprocal relationships.

\subsection{Analysis of Cyborgs Discussions}
A critical part of StratCom analysis is examining how an agent presents itself as a social actor in the creation of public culture and discussion of public issues \citep{hallahan2007defining}. We perform this analysis in two folds: stance detection analysis and topic analysis. Stance detection analysis indicates the alignment of the agents towards an issue. Topic analysis examines the discourse that agents aligning with each stance express.

We defined sets of hashtags related to pro/anti-vaccine, and conservative/liberals for the Coronavirus and US Elections events respectively. These sets were defined through manual inspection and classification of all the hashtags that had at least 10 occurrences in each dataset. The hashtag list was also pruned for the common and overly used hashtags like \#covidvaccine and \#vaccine that do not represent a stance. The full list of hashtags that are used in stance identification is available in the Supplementary Material. Having identified topics into opposing stances (pro- vs anti-vaccine in the Coronavirus dataset; conservative vs liberal in the US Elections dataset), we then analyzed whether Cyborgs and non-Cyborgs approach these topics differently.

We applied a stance propagation algorithm \citep{kumar2020social} to assign a stance to each agent using the structure of the all-communication network. This algorithm constructs a user-hashtag bipartite graph and propagate the stance labels between the user and hashtag portions iteratively. The algorithm returns a label (i.e. pro-, anti-, neutral) for each post and agent. For the purposes of this study, we examined data in the two opposing stances, pro- and anti, disregarding the agents that do not take a stand.

The assignment of stance to each agent splits the dataset of agents into two separate groups of conversations within the discourse of the event. With that segregation, we analyze the differences in topics discussed for Cyborgs and non-Cyborgs within each stance group, to examine whether Cyborgs post different types of messages as compared to non-Cyborgs. To analyze the topics discussed by each agent type, we used the topic modeling technique of Latent Dirichlet Allocation (LDA) on each set of agents respective to both events. LDA is a probability-based topic discovery algorithm that iteratively assigns topics to each through overall word distributions \citep{blei2003latent}. We used the Python implementation in the Gensim package\footnote{\url{https://radimrehurek.com/gensim/models/ldamodel.html}} to discover topics within each group. This algorithm returns a list of key terms that represent the top 5 topics discussed within the tweets, and manually combined the topics after inspection.

\subsection{Analysis of Cyborg profiles}
Finally, to answer the last research question on who Cyborgs agents are, we revisited the Cyborg profiles a year later. Some profiles were since suspended by the Twitter platform. Of the remaining alive profiles, we sampled an approximately 1\% subset (N=2857 for Coronavirus, N=1600 for elections) and manually categorized them to discover the nature of the accounts. Two of the authors scanned through the Cyborg agents and determined potential classes of agents. These proposed classes were discussed and harmonized to determine the final set of labels: renowned people and activists. Renowned people include persons of office, celebrities, and well-known people. Activists are people who advocate for political or social campaigns or changes. Following these definitions, the two authors independently assigned labels to each agent. In the case of disagreement, a third acted as a tie-breaker to determine the agent's label. Details about the annotation process are in the Supplementary Materials.

We also analyze the longevity of users in terms of three classes -- bot/cyborg/human -- hoping to understand the lifespan of the profiles in terms of their online behavior. We analyze the proportion of users that were suspended in each class a year later. Since we are unable to identify the exact date the users are suspended, we also make use of the still-alive users and analyze the length of time they are alive (number of days between the date of analysis from the date the user was created). Using data on the length of time the users are alive, we performed an ANOVA test across these three user classes, segregated by dataset. We also fit a linear regression line across the means of the three classes to visualize the trends in number of days alive.

\section{Results}
Cyborgs are defined as social media agents that are sometimes classified as a bot and sometimes classified as a human. That means, they often alternate between both bot and human classification and have excessive and extensive changes in their bot classification. These criteria characterize agents that are very likely bots in one stage and very likely humans in another timeframe. 

\subsection{Establishing a quantitative threshold for Cyborg classification}
We define Cyborg accounts as agents that (1) excessively flip bot classification and (2) have a large change in bot probability score between the flips. We first begin by deriving a quantitative criteria for Cyborg classification empirically. 
Through parsing each agent's bot classification day by day, we are able to establish quantitative thresholds for Cyborg classifications. 

\begin{figure}%[tbhp]
\centering
\includegraphics[width=\linewidth]{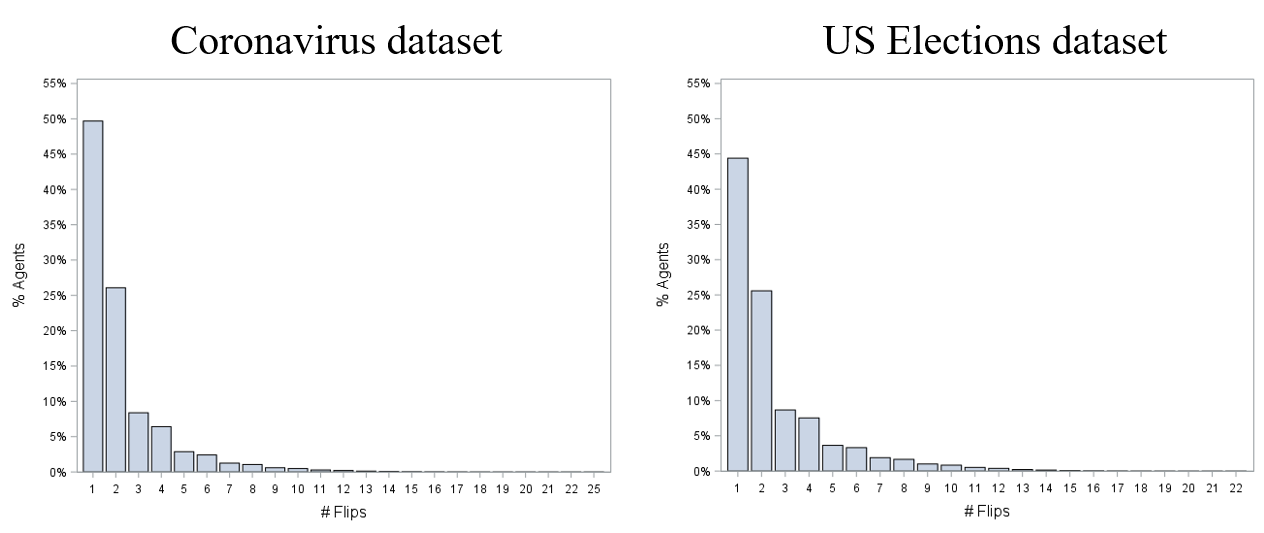}
\caption{Proportion of number of agents against number of flips within a month. Cyborgs are agents that flip more than thrice, which are the 75th-percentile of agents.}
\label{fig:numflips}
\end{figure}

\autoref{fig:numflips} shows the average proportion of agents against the number of flips, aggregating the data from the sub-datasets.  A large proportion of agents do not change its classification, lending weight to the bot classification model. We quantify excessive changes in bot classification as agents that change their classification more than thrice within the month. The number of flips at its 75th-percentile of agents is 3, which is observed against both datasets, locking in the value for our first Cyborg identification criterion. We selected the 75th-percentile through a sensitivity analysis of proportion of the agents that have at least $n$ number of flips. \autoref{fig:numflips} shows an exponentially decreasing proportion of agents as the number of flips increases, and we present the first few increasing number of flips in \autoref{tab:percentile}. We find that in the data for both events after 3 flips, the proportion of agents that flip $n$ times is less than 5\% and decreases in an exponential fashion, which is extremely small. Therefore, we select $n=3$ flips as a threshold, because beyond which, the proportion of agents starts to taper off. The upper limit of the proportion of agents with at least 3 flips is the 75th-percentile.

\begin{table*}%[tbhp]
\centering
\caption{Proportion of agents against number of flips of bot classification}
\label{tab:percentile}
\begin{tabular}{rrr}
Num of Flips & Coronavirus Dataset & US Elections Dataset \\ \hline
1 & 49.68 & 44.39 \\
$\leq$2 & 75.76 & 69.98\\ 
$\leq$3 & 84.14 & 78.64 \\ 
$\leq$4 & 90.56 & 86.18 \\ 
\bottomrule
\end{tabular}
\end{table*}

\begin{figure}%[tbhp]
\centering
\includegraphics[width=\linewidth]{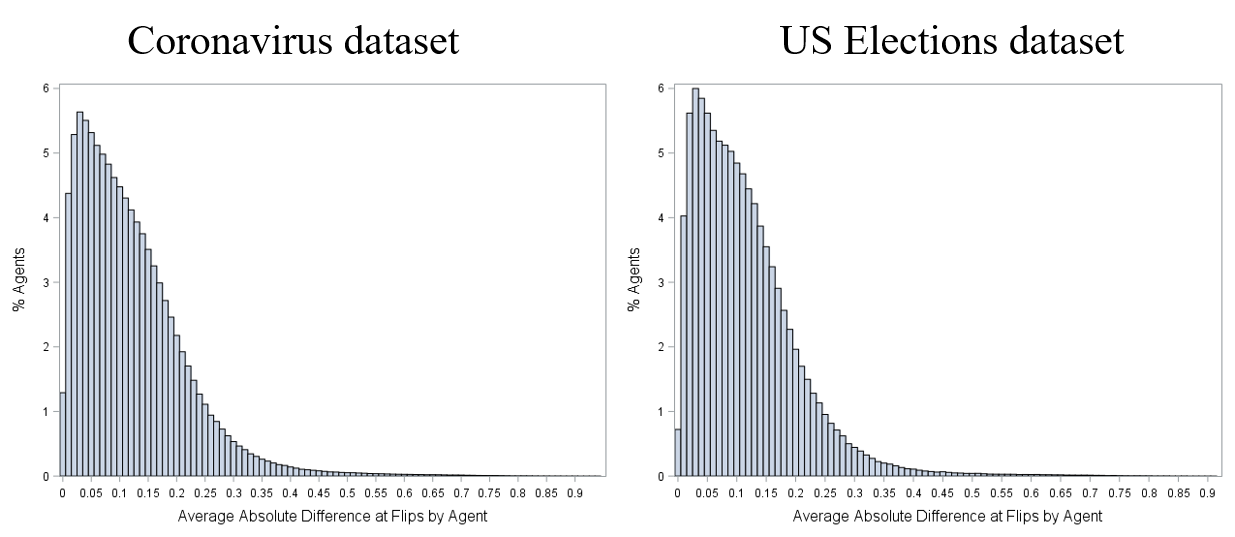}
\caption{Proportion of the number of agents against the absolute difference of bot probability score when the bot classification flips. Cyborgs are agents that flip bot classification with more than 0.10 score differences.}
\label{fig:absolutediff}
\end{figure}

\autoref{fig:absolutediff} shows the average proportion of the number of agents against the absolute difference of bot probability score when the bot classification flips. Both datasets present a mode of absolute difference of 0.05, which is usually not sufficient to change the bot classification of an agent.  
With this data aggregated across the sub-datasets, we observe that the average score difference at the 75th-percentile of agents is 0.10, which we use as a definition for our second Cyborg identification criterion. With such a huge change, the bot classification of the agent typically flips, and does so definitively (i.e., not hovering on the bot/human threshold border). To ensure external validity, a sample of these agents that are classified as Cyborgs are manually checked for their Cyborg nature by two of the authors who agreed on the Cyborg nature. This same sample is also used for categorizing the nature of the Cyborg accounts. More details on the sampling method and size are in the section ``Nature of Cyborg Accounts".

Among the Cyborgs, there is no significant differences between the directionality of bot classification flip. As detailed in \autoref{tab:flip_metrics}, there is an almost equal number of flips from bot to human as compared to human to bot accounts. This lends weight to the first definition of Cyborgs: that Cyborg accounts constantly flip between bot and human classification. Having an equal amount of flips between both directions shows that accounts do constantly change their behavior, and neither side of the duality is prominently favored over the other.

\begin{table*}%[tbhp]
\centering
\caption{Comparison of average number of times Cyborgs flip classification in the directionality of bot to human and human to bot}
\label{tab:flip_metrics}
\begin{tabular}{lrlr}
\multicolumn{2}{r}{Coronavirus Dataset} & \multicolumn{2}{r}{US Elections Dataset} \\
Bot to Human & Human to Bot & Bot to Human & Human to Bot \\ 
\midrule
3.59$\pm$3.35 & 3.61$\pm$3.35 & 4.04$\pm$3.65 & 4.09$\pm$3.66 \\
\bottomrule
\end{tabular}
\end{table*}

\begin{figure}%[tbhp]
\centering
\includegraphics[width=\linewidth]{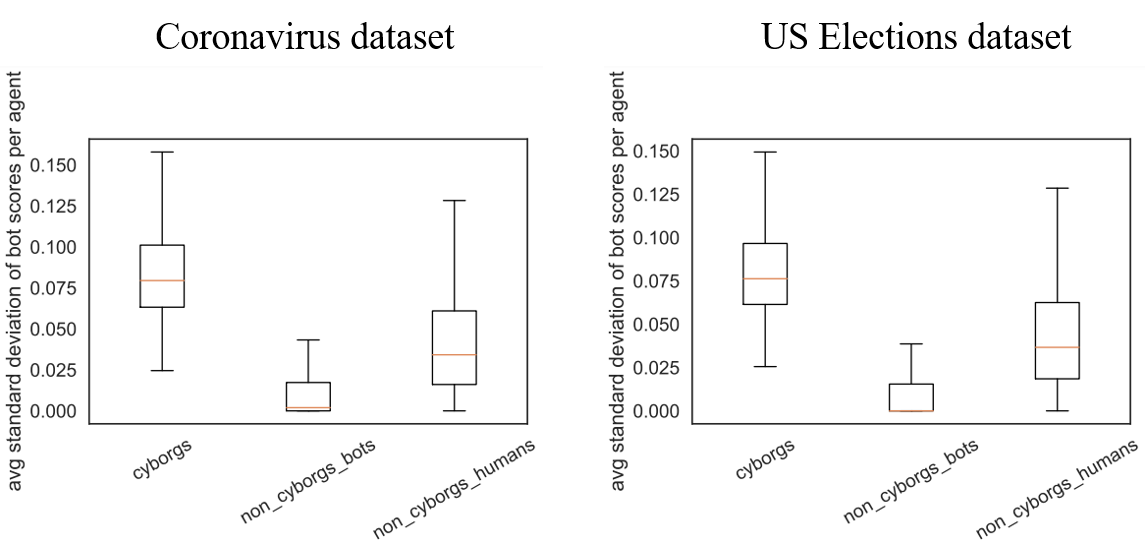}
\caption{Standard deviations of bot scores per type of agent.}
\label{fig:standarddeviation}
\end{figure}

Cyborgs also have a higher standard deviation in terms of their bot probability score, which is evidenced in their high frequency of changing bot classification (showcased in \autoref{fig:standarddeviation}). In non-Cyborgs, bots have a much lower standard deviation of bot probability score compared to humans, an observation that shows that bots are social media agents that have clearly defined features, likely because they are deliberately constructed.

\subsection{Network Centrality Analysis}
\autoref{tab:cyborgs_metrics} compresses the information for visual purposes, showing an X in the box where the metric is significantly higher. The full details of the network centrality analysis can be found in the Supplementary Material. The BotHunter algorithm does not use network properties to identify bots, hence a quantitative evaluation of network centrality provides insight towards the positions of Cyborgs/Non-Cyborgs within a communication network.

None of the Cyborg agents are verified Twitter accounts, but they have a higher number of followers and friends. A Twitter account is considered verified if the ``is\_verified" flag from the collected data is True. This corresponds to the blue checkmark on the Twitter web profile page. Cyborg agents are also centrally placed within a network: more connections as reflected by higher degree centrality scores, better positioned along the shortest path between other users as reflected by high betweenness centrality. Their eigenvector centrality scores are lower than non-Cyborgs, but that does not necessarily mean their influence is reduced; they themselves are influential nodes that other agents connect with, in hopes to activate their offline social influence, i.e. other agents may persuade a political Cyborg to increase vaccination sites in hopes that the human politician behind the account actually takes the corresponding action.

\begin{table}%[tbhp]
\centering
\caption{Comparison of metrics of Cyborgs and Non-Cyborgs agent classes. X marks the class with the significantly higher metric (p$<$0.001). }
\label{tab:cyborgs_metrics}
\begin{tabular}{lrr}
Metric & Cyborgs & Non-Cyborgs \\ 
\midrule
1. \% verified accounts & & X \\ 
2. Avg \# retweets & X & \\ 
3. Avg \# followers & X & \\ 
4. Avg \# friends & X & \\ 
5. Betweenness centrality & X & \\ 
6. Eigenvector centrality & & X \\ 
7. Degree centrality & X & \\ 
\bottomrule
\end{tabular}
\end{table}

\subsection{Topic Analysis}
We analyzed the discussions of agents of opposing stances in controversial issues to see if there is any differences in the topics that Cyborgs and non-Cyborgs post.
\autoref{tab:topics_coronavirus} and \autoref{tab:topics_elections} shows the topics retrieved by the topic modeling algorithm for each sub group. 
We observe that Cyborgs are present in all stance classes, highlighting that Cyborgs are used for strategic communications by all sides of the debate. In that aspect, there are no key differences in topical discussions between Cyborgs and non-Cyborgs within each stance group, indicating that the automation technology supports humans in information dissemination, most likely in message amplification through repeated retweets. In both events datasets, Cyborgs and non-Cyborgs alike carry out information campaigns to promote their ideologies and also take part in general online conversations. yborgs are also observed to be active on both sides of the debate, suggesting that these agents are harnessed for all types of uses, whether positive of negative.

\begin{table*}%[tbhp]
\centering
\caption{Topics discussed within the Coronavirus dataset}
\label{tab:topics_coronavirus}
\begin{tabular}{lp{7cm}p{7cm}}
& Pro-Vaccine & Anti-Vaccine \\ 
\midrule
Cyborgs & healthytogether, vaccineswork, pandemic, community, immunisation, globalhealth, callyourpediatrician, protect, ivax2protect, dangerous & vaccine, billgates, lies, fauci, facilities, covid19, vaccinedoesntwork, positive, mortality, takeresponsibility \\ 
Non-Cyborgs & coronavirus, covid19, testing, mask, deaths, breaking, record, pandemic, virus, spread  & nomask, novaccine, firefauci, billgates, endtheshutdown, endthelockdown, bigpharmaiskillingus \\ 
\bottomrule
\end{tabular}
\end{table*}

\begin{table*}%[tbhp]
\centering
\caption{Topics discussed within the US Elections dataset}
\label{tab:topics_elections}
\begin{tabular}{lp{7cm}p{7cm}}
& Conservative & Liberal \\ 
\midrule
Cyborgs & voting party, democrats, soldiers, putin, russia, projectlincoln, choice, racist, health, leadership, american, joebiden, gop, white, speakerpelosi & black, left, blm, statues, voter, mail, democrat, russia, taketheoath, justorder, child, border, executive, wwg1wga, potus, sidneypowell, taketheoath, qanon, tomfitton, dbongino, hejtlewis, cpompeo, genflynn, teamtrump, seanhannity\\ 
Non-Cyborgs & realdonaldtrump, tedlieu, markmeadows, racist, vote, projectlincoln, decent people wins, choice & berniesanders, black, obama, joebiden, maga, flynn, wwg1wga, unity, qanon\\ 
\bottomrule
\end{tabular}
\end{table*}

\subsection{Nature of Cyborg Accounts}
In a subsequent revisit of the Cyborg agents a year later, we find 23.5\% of Cyborgs suspended. Through a manual annotation, we find the following statistics in the Cyborg agents: 25\% were suspended, 36\% are activists, 27\% are renowned people (i.e. politicians, celebrities) and 12\% are other types of accounts (i.e. marketing, product communication, not StratCom etc). The results of the annotations are summarized in \autoref{tab:cyborgs_annotations}. The inter-annotator agreement (Cohen Kappa) of the initial two annotators that sorted agents into these categories is 0.87. This value is of the range [0.81,1.00] which represents ``almost perfect agreement", indicating that the Cyborgs can be distinctly differentiated into Activists, Renowned people and others \citep{mchugh2012interrater}. The annotations revealed that are in line with our hypotheses that majority of Cyborg accounts (63\%) are being used for communication purposes. Out of these 63\%, half (57\%) are used for activism, to promote topics through strategic messages and positioning within the network; and half (43\%) are used to provide temporary relief to the user in conveying important information.

\begin{table}%[tbhp]
\centering
\caption{Results of manual annotation of Cyborg agents}
\label{tab:cyborgs_annotations}
\begin{tabular}{rr}
Nature of agent & Proportion (\%) \\ 
\midrule
Suspended & 25 \\ 
Activists & 36 \\ 
Renowned Personalities & 27 \\
Others & 12 \\ 
\bottomrule
\end{tabular}
\end{table}

In terms of the lifespan of the users, \autoref{tab:longetivity} shows the longevity of the accounts. Across both datasets, we observe that the proportion of users suspended after a year is highest for bot users and lowest for human users. As for Cyborg users, there is a 50\% suspension rate. We further observe that Cyborgs have the highest average account length, followed by humans then bots, across both datasets. The comparison of the results of account lengths is statistically significant (p$<$0.05). This is also visualized in a graphical form, where a linear regression line formed by the means of the account lengths decreases from cyborgs to humans to bots.

\begin{table*}\centering
\caption{Comparison of proportion of users suspended and lifespan of still-alive users}
\label{tab:longetivity}
\begin{tabular}{p{3.5cm}rrrrrr}
 & Coronavirus dataset & & & US Elections dataset & & \\
& Bot & Cyborg & Human & Bot & Cyborg & Human \\ 
\midrule
Proportion suspended (\%) & 89.2 & 56.0 & 19.5 & 76.9 & 49.2 & 19.9\\
Avg length of acct (days) & 2751$\pm$1226 & 3663$\pm$1141 & 2901$\pm$1294 & 2643$\pm$1304 & 3437$\pm$1173 & 2715$\pm$1255 \\
p-value of ANOVA for length of acct& \multicolumn{3}{r}{3.452E-5} & \multicolumn{3}{r}{3.79E-142} \\ \hline
Graph of length of accounts against user type & \multicolumn{3}{r}{\includegraphics[width=0.4\textwidth]{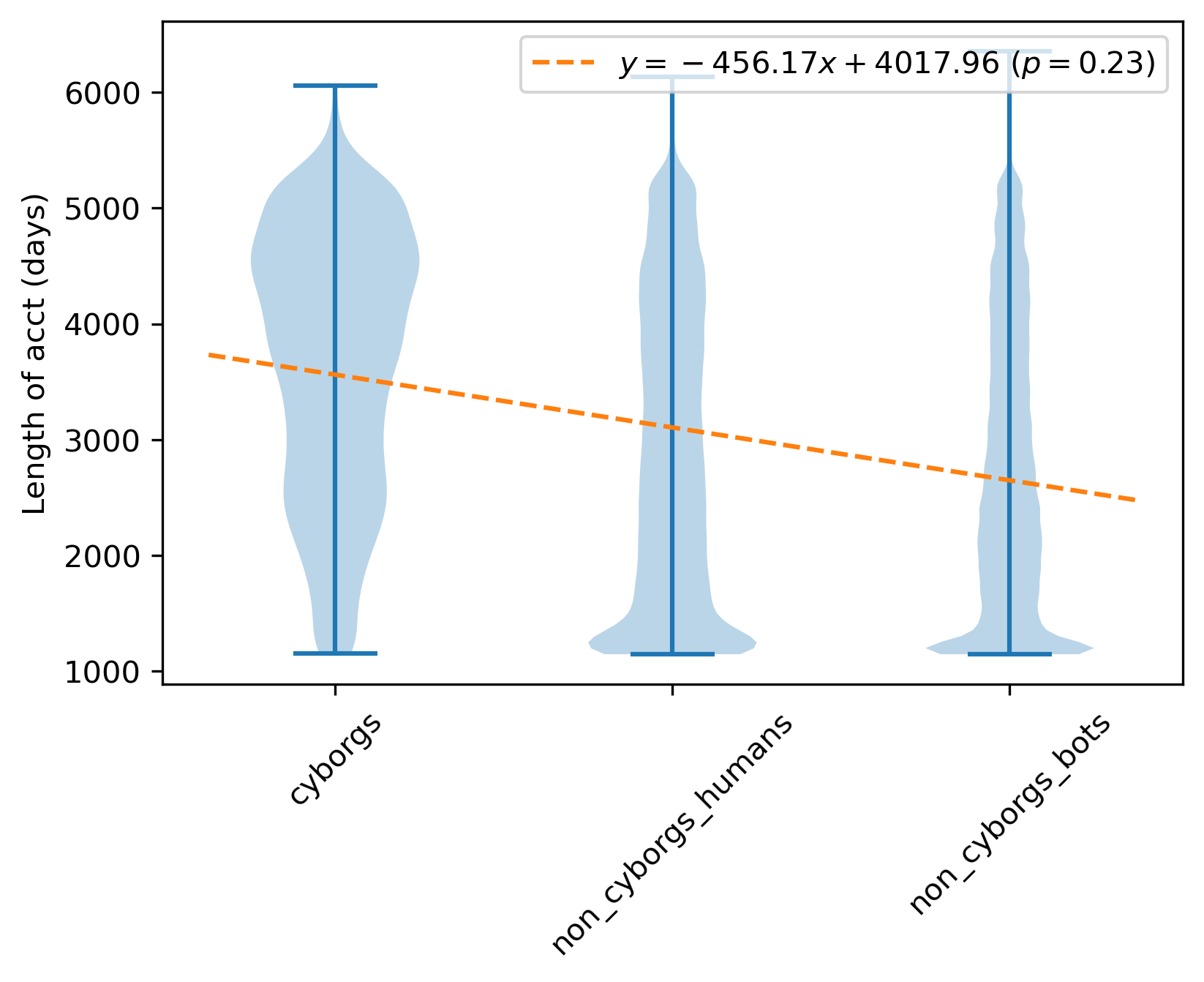}} & \multicolumn{3}{r}{\includegraphics[width=0.4\textwidth]{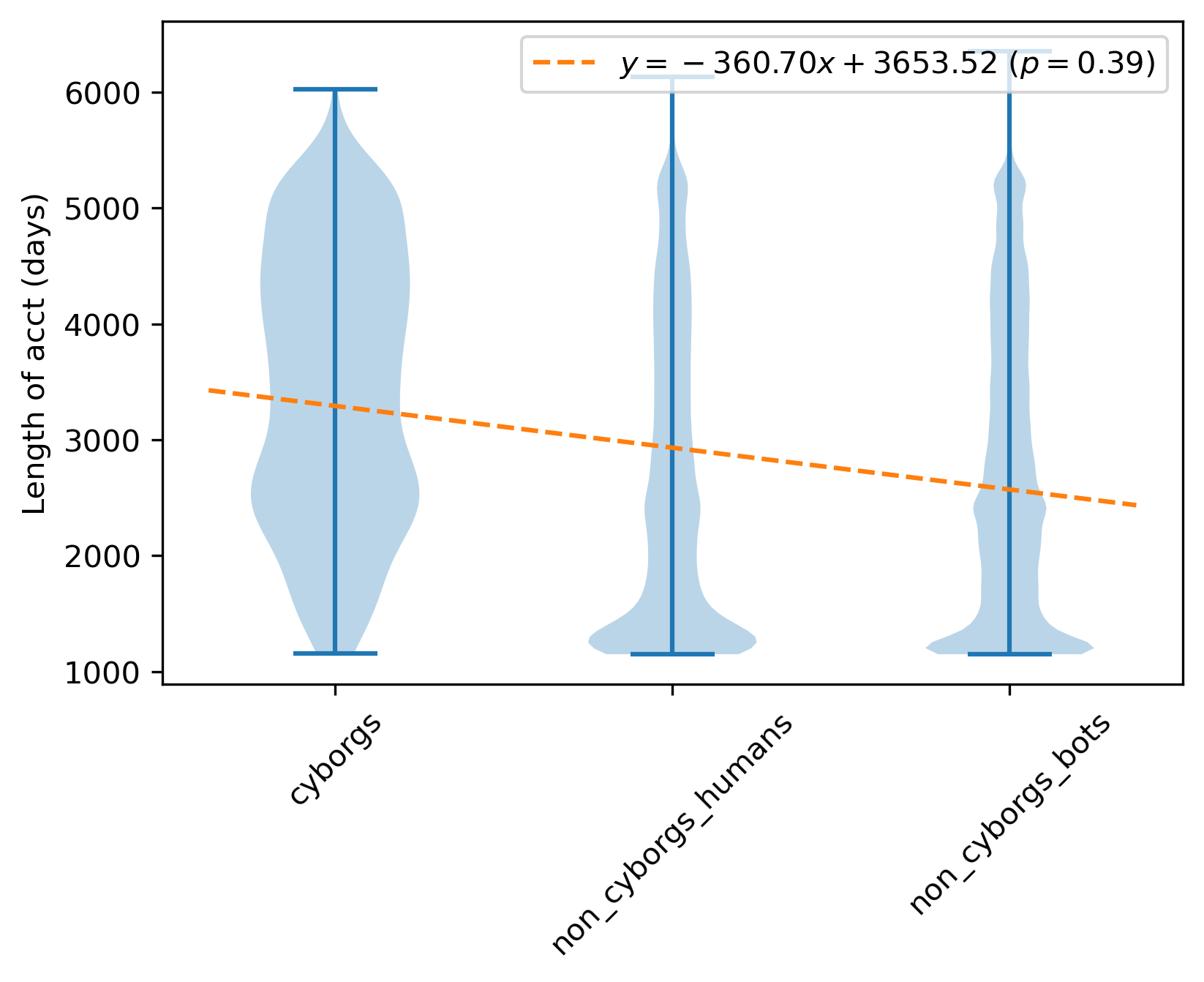}} \\
\bottomrule
\end{tabular}
\end{table*}

\section{Discussion}
Social media can be harnessed as an instrument for strategic communications. This platform is cleverly utilized by Cyborg agents, which are social media agents that sometimes appear to look like bot accounts, and other times appear as human accounts, to diffuse desired information about their personal or organizational brand. Here, we provided a data-driven definition of Cyborgs based on its fundamental characteristic of excessive and extensive bot classification flips: a Cyborg is a social media agent that flips bot classification more than thrice with an absolute bot probability score difference of at least 0.10 within a month. 

From the network centrality analysis, we discover that none of the Cyborgs are verified Twitter accounts. This indicates that despite some of these accounts being renowned people and activists, the platform does not confirm that their identity matches, likely due to the observation of bot-like behavior alongside human-like behavior; or that the account holders did not go through the verification process. Cyborg accounts spot high centrality values, suggesting that they are well-positioned within the all-communication network , giving them greater influence to connect and communicate with other users. They are information brokers, lying between groups of social media accounts, transporting and altering information through the system. They are well-connected, broadcasting desired information to a large number of accounts in the network. They have a larger number of friends and followers compared to non-Cyborgs, providing them a wide and vast reach of accounts that read the information they put out. These factors combined, identifies that Cyborg accounts are strategically positioned for communication and have sufficient influence to widely disseminate information.

Our annotations revealed that Cyborg accounts are typically used by Activists and Renowned Personalities, people who are trying to convey a specific message through the social media broadcast, and thus are strategic in who they broadcast to and how they broadcast it (as revealed in their network centrality and topic communication). This trends has also been observed in the 2017 UK general election, where human users volunteer their real profiles to be automated for political broadcasts or automated processes manned by human clickworkers to spread political content \citep{gorwa2020unpacking}.

Activists that are detected as Cyborgs generally have been around for at least five years, with the earliest identified being created in 2009. These Cyborgs tweet/retweet highly topical, mostly national and political issues, to generate attention. Examples are a staunch Republican conservative, who likely uses automation to disseminate information in support of that viewpoint, and sprinkles his opinions using the quote tweet feature. On the opposite end of the spectrum is a liberal user who likely uses automation to retweet stories in favor of the current Democratic president Biden, and also tweets his own personal evaluations of the situations (``Right wing media will be regurgitating gain of function for the next 21 days").
One Cyborg in the category of renowned people governs a UK country, uses a mixture of automation to retweet information regarding services and events in her county of rule, such as information on vaccination centers; and the human touch to broadcast personal events (``[...] I'm sure Nellie really appreciated your visit [...]"). Another Cyborg of a renowned person is a former UK Labor Parlimentary Candidate. She retweets information about the UK government policies. However, she is very much human in her other personal life-revealing tweets (``[...] Time to sort the veg for tomorrow and then Die Hard"; ``Happy Winter Solstice. After today the light comes back again."). 
An example of accounts classified in the ``Others" category is an entrepreneur who markets wealth planning expertise for business owners. This account frequently retweets financial market information and charts, and sometimes adds a touch of opinion through quote tweets. 

These accounts may not be entirely using automation in the bot-like phase, but are instead retweets by social media managers, these tweets appear bot-like and therefore bot detectors classify them as bots during those time periods. At other time periods, they are classified as human accounts. With the presence of excessive flipping bot classification, these accounts are classified as Cyborg for the duality of their appearance. For privacy reasons, we do not disclose the identifiers of the specific individuals.

The keywords obtained by the topic analysis module identifies pertinent discussion topics by Cyborgs and Non-Cyborgs. During the Coronavirus pandemic, vaccination was a hotly debated topic as authorities sought to the vaccine as a way to help lift the global disruption and cure the virus. While the pro-vaccination camp encouraged immunization through campaigns like ``healthy together" and ``vaccines work"; whereas non-Cyborgs were more concerned with the record number of deaths and virus testing requirements. The anti-vaccination camp shows campaigns from both classes of agents: conspiracy theories about Bill Gates planning to use the vaccines to implant surveillance microchips into people by Cyborgs and calls for the government to ``end the lockdown".

In the US Elections, we divided the stances of agents into the two dominant political leanings within the US: the Conservatives and Liberals. Conservatives Cyborgs posted topics relating to their presidential candidate Joe Biden and their campaign promises such as health bills; and non-Cyborgs posted similar topics. Liberal Cyborgs and non-Cyborgs latched on to several campaigns such as Black Live Matters (``\#blm") and QAnon campaigns (i.e., ``\#wwg1wga", ``\#qanon"). Liberal Cyborgs have a larger range of topics that they campaigned on Twitter as compared to the non-Cyborg group. 

Cyborgs being active on both sides of a debate shows that these agents are harnessed for all types of uses, and are not deployed only for malicious use. They are observed on all sides to be spreading narratives in support of particular campaigns, supporting their use for strategic communications on the social media sphere. Some Cyborg agents from the coronavirus dataset have affiliations to health authorities and are used for promoting mass vaccination among the public, frequently posting about the advantages of vaccination. On the other hand, anti-vaccination Cyborgs spread the message that ``vaccines don't work" which increases vaccine avoidance. Cyborgs are also used for political purposes, as observed in the US Elections dataset. They are used to promote the campaigns of political parties, and also of their ideologies and slogans (i.e. ``taketheoath"). Given that Cyborgs and Non-Cyborgs discuss similar topic sets, and promote ideologies, what differentiates Cyborgs and non-Cyborgs, then are not the discussion of issues but \textit{who} the agents are. We observe this through the network and profiling analysis: Cyborgs are more influential agents within its communication social network, and the nature of agents are activists and renowned personalities.

Among the entire dataset, only 23.5\% of the Cyborgs are suspended, which is lower than the proportion of agents being suspended in other bot detection studies. These studies involve general collection and US elections revealed 30 to 99.4\% of the collected and identified bot accounts being subsequently suspended \citep{ferrara2020bots,volkova2017identifying}. From studying the proportion of suspensions and the average account length, we note that Cyborgs have the longest average lifespan, and are less likely to be suspended as compared to bot accounts. The longevity of Cyborg accounts point to their effectiveness at evading bot detection classifiers and anti-spam filters (which have also been observed in previous studies \citep{grimme2018changing,gorwa2020unpacking}), or the graciousness of the social media platforms to allow them to continue their operations. Their longer lifespan, despite being unverified accounts, make them more alluring as StratCom devices, as it reduces the need for their human operators to constantly create new social media presences to market their brand yet provides an account that endures for a long time, sometimes years. 

In our studies, we find similar observations within the two datasets (Coronavirus and US Elections), indicating applicability of our observations to a the wider social media ecosystem. Similar characteristics of Cyborg accounts across two large scale datasets point to generalizability of the Cyborg behavior on social media. This sets about thresholds corresponding to the definitions of Cyborgs, in particular the number of flips and the average difference in bot score during flips. This work contributes to the study of automated account detection on social media, to provide a repeatable method for identifying Cyborg accounts that throw off current machine-learning-based bot detection models based on feature set matching with the combination of human engagement and automated participation.

\subsection{Study Limitations and Future Research}
Although ours is among the largest study of Cyborgs on social media to date, we caution that there is a wide range of users on social media and future studies are needed to further discover and characterize the nature and intent behind hybrid accounts. While our observations identify that most Cyborg accounts are used for StratCom purposes, there may be a better separation of the intention and type of communication put forth, that requires deeper analysis of each agent's profile.

The identification of Cyborgs requires a longitudinal study, which is largely dependent on the data collected through the provided Twitter API. To successfully perform this study, one requires the posts of each agent across a sufficiently long period of time, which may not always be available due to data collection limitations. Despite identifying values, we note that with this method, in order to successfully identify Cyborgs within a dataset, one must have sufficient longitudinal data about the users; too short a timeframe and no flipping behavior is observable. We used a monthly timeframe for our experiments, a value that can be adapted to ensure sufficient data accumulation.

Nonetheless, we hope our large scale statistical analysis aids in determining the amount of automation a human agent needs to exhibit; or the amount of human touch an automated bot agent needs to possess to be classified as a Cyborg \citep{gorwa2020unpacking}. Future work includes in-depth analysis of Cyborgs using network science approaches across different domains and social media platforms, providing detailed characterizations to their online behavior.

\section{Conclusion}
Given that automated social media agents can inorganically disrupt the online discourse \citep{ferrara2020bots}, it is important to contextualize the role of Cyborg agents within the ecosystem. In our experiments, we utilized the BotHunter bot detection algorithm to classify agents across time, and hence summarize our recommendations for parameters as threshold values to identify Cyborgs: a social media agent with at least (1) 3 flips of bot classification, from bot to human or human to bot; and (2) the average change in bot probability score between the flips is at least 0.10. 

Deeper analysis of Cyborg accounts indicate that Cyborgs are used as strategic communication devices for renowned personalities and activists, and are used by all factions of a debate: there are good Cyborgs as there are bad Cyborgs. Cyborg agents appear both has bots and humans in their information dissemination behavior, and are typically well-positioned within the social network, which increases the reach of their constructed information. 

This study used a mixed-methods approach to analyzing Cyborgs. From a quantitative point of view, it provided numerical thresholds to classifying Cyborg accounts, derived from observations of two large datasets. This provides a repeatable method for identifying Cyborg accounts. From a qualitative point of view, it characterized the type of accounts Cyborgs present in terms of renowned personalities, activists and others.  We hope that this study has shed light into one of the personalities on social media - the hybrid persona between bot and human.

\section{Declaration of conflicting interests}
The authors declare no potential conflicts of interest with respect to the research, authorship and/or publication of this article.

\section{Data Availability}
The data analyzed in this study are not publicly available due to the risk of inadvertent disclosure of identifiers of specific individuals or entities. However, the data may be available on reasonable request, in accordance to Twitter’s Terms and Conditions.

\begin{acks}
%Removed for Anonymity
This research was supported in part by the following organizations and grants: the Knight Foundation, the Office of Naval Research (Bot Hunter, Grant N000141812108, Group Polarization in Social Media N000141812106), the Center for Computational Analysis of Social and Organizational Systems (CASOS), the Center for Informed Democracy and Social Cybersecurity (IDeaS) at Carnegie Mellon University. The views and conclusions contained in this document are those of the authors and should not be interpreted as representing the official policies, either expressed or implied, of Knight Foundation, the Office of Naval Research or the U.S. Government.
\end{acks}

\bibliographystyle{SageH}
\bibliography{main}

\clearpage

\appendix
\section{Supplementary Material}
\subsection{Hashtags used for data collection}
We selected the following hashtags for the US Elections and Coronavirus events for data collection. We tried to keep the terms as neutral as possible, and include terms from both sides of the event (e.g. Democrats and Republicans). These hashtags were identified from the tweets themselves using the property of co-occuring hashtags during the data collection.

\underline{Coronavirus dataset:} \\
The hashtags used for data collection of the Coronavirus dataset are: \#coronavirus, \#coronaravirus, \#wuhanvirus, \#2019nCoV, \#NCoV, \#NCoV2019, \#covid-19, \#covid19.

The data analyzed in the study came from the following months: June 2020, August 2020, November 2020, January 2021, February 2021 and May 2021.

\underline{US Elections dataset:} \\
The hashtags used for data collection of the 2020 US Elections dataset are: \#election2020, \#2020\_presidential\_election, \#maga2020, \#flipitblue, \#keepitblue, \#yeswecan, \#yang2020, \#JoeBiden, \#BernieSanders, \#ElizabethWarren, \#PeteButtigieg, \#FeelTheBern, \#democrats, \#republicans

The data analyzed in the study came from the following months: April 2020, June 2020, August 2020, November 2020, January 2021 and February 2021.

\subsection{Hashtags used for Stance Detection}
The hashtags used for the Coronavirus datasets are:
\begin{enumerate}
    \item Pro-Vaccine: VaccinesWork, Sharethevaccine, ProtectVaccineProgress, getvaccine, WaitforVaccine, FreeTheVaccine, vaccinesaresafe, vaccineconfidence, igotvaccinated, coronavaccineforall, CoronavirusVaccineAppointment, justtakethefuckingvaccine, VaccinesWithoutBorders, Vaccines4All, vaccinesaves, Vaccine4ALL, BreakthroughVaccine, waitingformyvaccine, GetTheVaccine, GetYourVaccine, takethevaccine, vaccinesafe, Vaccineswillwork, Iwilltakethevaccine, vaccinefreedom, VaccineToAll, SafeAndEffectiveCovid19Vaccine, VACCINESARESAFE, safevaccines, nosleeptilvaccine, NoOnsiteSchoolsUntilVACCINES, WhereIsMyVaccine, vaccineselfie, vaccineready, vaccinesareamazing, vaccinee, VirusToVaccine2020, HoHoHopeVaccineArrivesSoon, Vaccined, VaccinesWork, VaccinesSaveLives, Vaccine4All, effectivevaccine, VaccineForSA, VaccinesSavesLives, SafeVaccines, accesstovaccines, VaccineHope, VaccinesWithoutBorders, wherearethevaccines, VaccinesforAll, getthevaccine, giveusthevaccine, VaccineWorks, vaccinesavelifes, GetAVaccine, vaccinesafelife, safevaccine, HaveTheVaccine, WhyIGotMyVaccine, CovidVaccineforall, vaccinesavestheworld, ImGettingTheVaccine, FreeVaccines, VaccinesWorkforAll, GiveMeMyVaccineNow, Affordablevaccine4all, getthatvaccine, justgivemethevaccine, SayYestotheVaccine, TakeYourVaccine, provaccine, YesToVaccine, vaccinesave, covid19\_vaccine\_4all, vaccineissafe, PleaseGetTheVaccine, VaccinesForEveryone, TrustTheVaccine, vaccinessavelives, VaccinesWork, Vaccine4All, TakeTheVaccine, FirstDoseOfVaccine, We4Vaccine, vaccinesaveslives, VACCINEFORWELLNESS, TheVaccineIsSafe, IWillTakeTheVaccine, firstdosevaccine, IWillTakeVaccine, SafeVaccine, SayYesToVaccine, VaccinesWorkForAll, VaccineIsSafe, TrustTheVaccine, getyourvaccine, SafeVaccines, VaccineSavesLives, vaccinesaresafe, YesToCovid19Vaccine
    \item Anti-Vaccine: NoVaccines, NoVaccinesForMe, VaccineYourAss, NoToCoronavirusVaccines, SayNoToVaccine, NoMandatoryVaccine, VaccinesKill, VaccinesKills, stopvaccine, fkyourvaccines, antivaccine, AntiVaccine, NoVaccine, StopCovidVaccine, WeDoNotConsentCVVaccine, VaccinesKill, VaccinesHarm, SayNoToVaccines, ForcedVaccines, VaccineIsPoison, ResistVaccines, Noneedvaccines, FakeCoronaVaccine, VaccinesAreBioweapons, Iwontgetthevaccine, VaccineFromHell, vaccinepoison, StopAllVaccines, Vaccinetakedown, vaccinesDAMANGEimmunity, vaccinesRnotNATURAL, RejectTheVaccines, BewareVaccines, FuckVaccines, HellNoVaccine, NoVaccinesEver, WhoNeedsVaccine, justsaynotoforcedvaccines, StickTheVaccineUpYourArse, nottakingavaccine, vaccinebad, WeaponizedDeadlyVaccines, JustSayNOToTheVaccines, DoNotTakeTheVaccine, FuckVaccinePoison, NOVaccine4Me, VaccinesNotSafe, VaccineBioWeapon, wedontwantvaccine, vaccinenotforme, DontTakeCovidVaccine, NotTakingCovidVaccine, DangerousVaccine, vaccinedoesnotwork, VaccineIsntSafe, No2Vaccine, OpposeTheVaccine, YouCanHaveMyVaccine, ShoveThatVaccineUpYourAsshole, MoThankYouCovidVaccine, ToHellWithCovidVaccine, CovidVaccinePoison, SCREWTHEVACCINE, murdervaccine, VaccineIsUseless, KilloronaVaccine, DodgyVaccine, DoNotTakeAnyVaccine, JeNeMeVaccineraiPas, NoVaccineForMe, stopthevaccine, novaccine, BoycottIndianVaccine, vaccinehesitancy, To\_Vaccine\_Is\_My\_Choice, vaccinefailure, donttakethevaccine, FakeVaccine, StolenVaccines, NoVaccine4Me, TrudeauVaccineContractsLie, vaccinesKill, VaccinesArePoison, VaccinesAreNotCures, NotAVaccine, jesusisvaccine, HALTtheVaccines, NoVaccines, dontgetthevaccine, VaccinesHarm, TeamNoVaccine, TheVaccineIsTheVirus, fakeVaccines, killervaccine, lethalvaccine, CoronaVaccineFail, vaccineBioweapon, saynotovaccines, Fuckvaccines, fuckyourvaccines, vaccinedeath, JustSayMoToVaccines, NoCOVIDVAccineMandate, NoMandatoryVaccines, GoHomeLeaveOurVaccinesAlone, antivaccine, anti\_vaccine, stopcovidvaccine, CovidVaccineHesitancy, VaccinesCanKill, Antivaccines, notovaccine, DontGetVaccine, VaccineDontWork, VaccineKills, StopVaccine, ForcedVaccine, FakeCovidVaccine, VaccinesAreBad, Falsevaccine, PfizerVaccineKillingPeople, NoCovidVaccineForMe, VaccineShaming, TakeTheVaccineAndShoveItUpYourAss, FakeVaccinesWillNotSaveYou, WorldSaysNoVaccine, NOCovidVaccine, NOCovid19Vaccine, Jesusovervaccines, IAmNOTVAccineBait, NotAVaccineAMedicalExperiment, VaccineMortality, NoVaccine, NoVaccineForMe, NOVACCINE4ME, VaccineDeaths, VaccineDeath, VaccineInjury, AntiVaccine, vaccinesideeffect, NoToCoronaVirusVaccines, AvoidCovidVaccine, NoVaccines, coronavirusvaccinescam, ForcedVaccines, destroyvaccines, VaccineHesitancy, noneed4vaccine, notocovidvaccine, vaccinesharm, vaccinedamage, vaccinesareevil, SayNoToVaccines, killervaccine, no2vaccine, vaccinekills, vaccineskill, VaccinesKillingpeople, JustSayNotoVaccines, vaccinedanger, donttrustthecovid19vaccine, RejectWeaponizedVaccines, StopVaccine, FakeVaccine, vaccinechemicalweapon, DeathToVaccines, VaccineScam, NoVaccinesNeeded, NoVaccinesForMe, vaccinehesitant, PoisonVaccine, DeathVaccine, Cancervaccine, VaccineFail, vaccineRESISTANT, VaccineNonsense, UnsafeVaccines, NoVaccinesNecessary, AbnominableVACCINE, Vaccinefuckup, novaccinerequired, vaccinedrivemutations, VaccineFromHell, GodIsMyVaccine, AntiCovidVaccine, NeitherDoTheseCOVIDVaccines, notovaccine, Notmyvaccine, CovidVaccineIsPoison, VaccineDisaster, NovavaxVaccine, TheCovidVaccineKills, vaccinekills, StopTheVaccines, fuckyourvaccine, VaccineIssues, vaccinesDONTwork, vaccinebad, VaccineNotTheAnswer, dontgetthecovidvaccine, MurderbyVaccine, vaccineforwhat, norealvaccine, NoVaccine, VaccineScam, VaccineDeaths, NotAVaccine, VaccineKills, NOVACCINE4ME, vaccineisdeath, vaccineispoison, NoVaccineForMe, To\_Vaccine\_Is\_My\_Choice, vaccinesharm, NoVaccinePassports, NoToVaccinePassports, VaccinesAreDangerous, VaccinesKill, VaccinesAreBad, saynotovaccines, FuckTheVaccine, StillAintGettingTheVaccineThough, ChineseVaccineBioWeapon, antivaccine, JustSayNotoVaccines, Notomandatoryvaccines, fakevaccines, vaccinedeath , vaccineFail, StopVaccinePassports, CovidvaccineFail, COVID19VaccinesBioWeaponsOfMassDestruction, NoMandatoryVaccines, DoNotGetTheVaccine
\end{enumerate}

The hashtags used for the US Elections datasets are:
\begin{enumerate}
    \item Conservative: conservative, trump, republican, maga, politics, donaldtrump, makeamericagreatagain, kag, trumptrain, conservativememes, americafirst, keepamericagreat, republicans, gop, libertarian, conservatives, rightwing, prolife, draintheswamp, biden, democrats, trumpsupporters, liberty, patriot, capitalism, election, politicalmemes, trumpmemes, alllivesmatter, fakenews, presidenttrump, socialism, walkaway, redpill, blm, buildthewall, communism, republicanparty, potus, christian, constitution, conservativewomen, ndamendment, republicanmemes, bluelivesmatter, truth, redwave, socialismsucks, usa, truth, freedom, america, news, trump, christian, blm, politics, donaldtrump, maga, republican, conservative, makeamericagreatagain, alllivesmatter, conservative, blm, politics, donaldtrump, maga, liberty, republican, conservative, makeamericagreatagain, alllivesmatter, fakenews, patriot, bluelivesmatter, trumptrain, democrats, prolife, potus, conservative, americafirst, republicans, draintheswamp, walkaway, presidenttrump, kag, politicalmemes, buildthewall, redpill, qanon, keepamericagreat, rightwing, conservatives, trumpsupporters, conservativememes, redwave, republicanparty, republicanmemes, conservativewomen, ndamendment, conservative, conservatives, conservativememes, conservativewomen, conservativeandproud, conservativeparty, conservativemovement, conservativelogic, conservativenews, Conservativesatire, conservativeamerica, conservativememe, conservativepolitics, conservativelibertarian, conservativevalues, conservativefashion, conservativecomedy, conservativegirls, conservativessuck, Conservativepride, conservativemom
    \item Liberal: liberal, politics, maga, democrats, america, libertarian, donaldtrump, usa, freedom, socialism, election, capitalism, blm, joebiden, politicalmemes, makeamericagreatagain, democrat, liberalismo, communism, vote, republicans, blacklivesmatter, liberty, berniesanders, gop, kag, covid, trumptrain, voteblue, progressive, leftist, liberals, rightwing, americafirst, socialist, news, feminism, argentina, draintheswamp, bernie, resist, bidenharris, libertad, conservatives, political, capitalismo, prolife, feminist, lgbt, keepamericagreat, liberal, freedom, argentina, news, blacklivesmatter, trump, covid, feminism, vote, blm, politics, feminist, donaldtrump, libertad, liberty
\end{enumerate}

\pagebreak
\subsection{Elaboration on Identifying Cyborgs}
This section elaborates on the use of bot detection algorithms in identifying cyborgs, to supplement the description in the main manuscript.

There are many bot detection algorithms constructed to differentiate whether an agent is a bot or human, e.g. Botometer \citep{yang2020scalable}, BotHunter \citep{beskow2018bot}, DeBot \citep{chavoshi2016debot} etc. After evaluating the algorithms, we chose the BotHunter algorithm. BotHunter can use historical data to determine the classification of an agent, instead of relying on real-time data pulls like Botometer does, or matching with a curated set of identified agents like DeBot. The BotHunter algorithm returns a bot probability score in the range of 0 to 1, where 1 means that the agent is most likely a bot and 0 means that the agent is most likely a human. The BotHunter model is trained on publicly available Twitter datasets, some of which are used to train the Botometer model too. It classifies Twitter agents via supervised random forests in a multi-tiered approach. We use the Tier 2 model of BotHunter, which uses account features, account meta-data and temporal features to determine a bot classification. We use the Tier 2 of BotHunter, which uses account features, account meta-data and temporal features to determine a bot classification. We did not use network features within this evaluation because the data collection of the network features of almost 63 million agents is extremely time intensive.

To determine whether an agent is a bot or not, we make use of thresholding the bot probability score. In literature, many different values have been used: 0.25 \citep{zhang2019whose}, 0.50 \citep{boichak2018automated}, 0.70 \citep{rauchfleisch2020false} and so forth. We used the 0.70 threshold to classify an agent as a bot or not. That is, if an agent's bot probability is greater or equal to 0.70, it is classified as a bot; otherwise it is classified as a human. This threshold was chosen in reference to a systematic analysis of levels that give stable bot classification scores, whereby most agents do not change their bot class \citep{ng2022stabilizing}.

With the bot classification of each agent, we trace the number of times an agent flips bot classification across the months. To do this, we segment an agent's tweets into separate days and evaluate the agent's bot classification day by day, and compare it to the previous day. The agent is determined as to have flipped bot classification if the classification differs between the two time periods. BotHunter facilitates this quantitative evaluation because it allows reading in of tweets from a JSON file, meaning we can control the range and number of tweets ourselves. While the flipping behavior can be due to false positives, we control this by not taking the agents that flip only once or twice to be considered as Cyborgs. By using a threshold of Cyborgs as having more than 3 flips during our year-long data collection, we identify only 15.86\% of the agents as Cyborgs, which is a reasonable proportion. Given that the number of agents exponentially decreases with the increasing number of flips (Figure 1), a good proportion of this flipping behavior is not due to classifier error. Lastly, we analyze a large number of agents -- 63 million agents -- therefore, while there are false positives results, the large dataset should balance out the false classification.

\pagebreak
\subsection{Elaboration on Annotation Process}
We manually categorized the identified Cyborg agents to discover the nature of accounts. This process begins with constructing a dataset from the sample of identified Cyborg agents. We randomly select 1\% of agents from each event, resulting in a dataset of 2857 Cyborgs from the Coronavirus data, and 1600 Cyborgs from the US Elections data. The data points were shuffled to prevent any subconcious ordering. Two authors scanned through the dataset, which included account's user meta-data and tweets, and proposed categories to describe the nature of the accounts. After discussion, the categories were harmonized, resulting in three main categories: Activists, Renowned People, and Others; for the agents that appear to be Activists and Renowned People stood out distinctly.

After deciding on these categories, the two annotators annotated each agent and indicated whether each agent is an Activist or Renowned person or others. This annotation process achieved an inter-annotator agreement, measured by Cohen's Kappa, of 0.87 between the two initial annotators. The third annotator broke the differences from the initial annotation before the annotation set was finalized.

From this manual annotation process, we find the following statistics in the Cyborg agents: 25\% were suspended, 36\% are activists, 27\% are renowned people (i.e. politicians, celebrities) and 12\% are other types of accounts (i.e. marketing, product communication, not StratCom etc). Unfortunately, because many of the accounts are still in operation on the social media platform, we refrain from providing identifying these users.

\pagebreak
\subsection{Network Centrality Analysis Results}
\begin{table*}%[tbhp]
\centering
\caption{Comparison of metrics of Cyborgs and Non-Cyborgs agent classes. * marks the class with the significantly higher metric (p$<$0.001). }
\label{tab:cyborgs_metrics_full}
\begin{tabular}{lrrrr}
& \multicolumn{2}{r}{Coronavirus Dataset} & \multicolumn{2}{r}{US Elections Dataset} \\
Metric & Cyborgs & Non-Cyborgs & Cyborgs & Non-Cyborgs \\ 
\midrule
1. \% verified accounts & 0 & 100* & 0 & 100* \\ 
2. Avg \# retweets & 46.25* & 23.24 & 70.35* & 60.84\\ 
3. Avg \# followers & 2001.11* & 1920.79 & 1690.65* & 1410.21\\ 
4. Avg \# friends & 1459.85* & 682.61 & 1899.64* & 744.07\\ 
5. Betweenness centrality & 9.79E-4* & 3.99E-4 & 7.55E-4* & 5.76E-4\\ 
6. Eigenvector centrality & 3.65E-6 & 2.12E-5* & 2.14E-4 & 1.72E-3*\\ 
7. Degree centrality & 8.54E-9* & 3.70E-11 & 5.40E-9* & 7.59E-11\\ 
\bottomrule
\end{tabular}
\end{table*}

\autoref{tab:cyborgs_metrics_full} showcases the comparison of metric values between Cyborgs and Non-Cyborg classes for both events. Significant testing is performed on the two series of metrics at the p$<$0.001 level, with * indicating the value that is significantly higher. \autoref{tab:cyborgs_metrics_combined} showcases the summarized comparison of metric values between Cyborgs and Non-Cyborg classes. It averages the metrics from both events in this comparison. 

\begin{table}%[tbhp]
\centering
\caption{Comparison of metrics of Cyborgs and Non-Cyborgs agent classes. * marks the class with the significantly higher metric (p$<$0.001). }
\label{tab:cyborgs_metrics_combined}
\begin{tabular}{lrr}
Metric & Cyborgs & Non-Cyborgs \\ 
\midrule
1. \% verified accounts & 0 & 100* \\ 
2. Avg \# retweets & 58.3* & 42.04\\ 
3. Avg \# followers & 1845.88* & 1665.5\\ 
4. Avg \# friends & 1679.75* & 713.34\\ 
5. Betweenness centrality & 8.67E-4* & 4.88E-4\\ 
6. Eigenvector centrality & 1.09E-4 & 8.71E-4* \\ 
7. Degree centrality & 6.97E-9* & 5.65E-11\\ 
\bottomrule
\end{tabular}
\end{table}

\end{document}